\documentclass[pra,twocolumn,superscriptaddress,showpacs,eqsecnum]{revtex4}
\usepackage{amsmath}
\usepackage{amsfonts}
\usepackage{graphicx}
\usepackage{color}

\begin{document}

\title{
        Unitary transformations for testing Bell inequalities
}
\author{S.\ D.\ Bartlett}
\affiliation{Department of Physics, Macquarie University, Sydney,
        New South Wales 2109, Australia}
\author{D.\ A.\ Rice}
\affiliation{Department of Physics, Macquarie University, Sydney,
        New South Wales 2109, Australia}
\author{B.\ C.\ Sanders}
\affiliation{Department of Physics, Macquarie University, Sydney,
        New South Wales 2109, Australia}
\affiliation{Erwin Schr\"odinger International Institute for
        Mathematical Physics, Boltzmanngasse 9, A--1090 Vienna, Austria}
\author{J.\ Daboul}
\affiliation{Department of Physics, Macquarie University, Sydney,
        New South Wales 2109, Australia}
\affiliation{Department of Physics, Ben-Gurion University of the Negev, 
        Beer--Sheva 84105, Israel}
\author{H.\ \surname{de Guise}}
\affiliation{Department of Physics, Macquarie University, Sydney,
        New South Wales 2109, Australia}
\affiliation{Centre de Recherche Math\'ematique, Universit\'e de
        Montr\'eal, C.P.\ 6128--A, Montr\'eal, Qu\'ebec H3C 3J7, Canada}
\date{January 8, 2001}
\begin{abstract}
  It is shown that optical experimental tests of Bell inequality
  violations can be described by SU(1,1) transformations of the vacuum
  state, followed by photon coincidence detections.  The set of all
  possible tests are described by various SU(1,1) subgroups of
  Sp(8,$\Bbb R$).  In addition to establishing a common formalism for
  physically distinct Bell inequality tests, the similarities and
  differences of post--selected tests of Bell inequality violations
  are also made clear.  A consequence of this analysis is that Bell
  inequality tests are performed on a very general version of SU(1,1)
  coherent states, and the theoretical violation of the Bell
  inequality by coincidence detection is calculated and discussed.
  This group theoretical approach to Bell states is relevant to Bell
  state measurements, which are performed, for example, in quantum
  teleportation.
\end{abstract}

\maketitle

\section{Introduction}
\label{sec:intro}

The controversy regarding the completeness of quantum
mechanics~\cite{Ein35} was presented in the framework of entangled
spin--$1/2$ particles~\cite{Boh51}.  This context proved to be
convenient for Bell's development of an inequality to test the
postulates of local realism~\cite{Bel64,Cla69,Cla78,Per93}.  Recent
quantum optics experiments, designed to test Bell inequalities,
involve pairs of photons that are produced from the vacuum state,
generally by optical parametric down-conversion (PDC).  PDC offers
significant advantages over the earlier atomic cascade approach to
generating photon pairs~\cite{Asp82}; these advantages include
conservation of energy (hence correlated frequencies of the two
photons), conservation of linear momentum (hence correlated
wavelengths and direction of propagation) and conservation of angular
momentum (hence correlated polarizations), as well as
near--simultaneity of the emission of the two photons in the
pair~\cite{Hon85}.  In addition to PDC acting as a source of
correlated pairs of photons, there exists a scheme for which the
photon pairs are in a polarization--entangled state~\cite{Kwi95}.  PDC
has enabled accurate tests of local realism vs quantum theory to be
performed~\cite{Kwi95,Ou88,Shi88,Rar90,Fra91,Bre92,Kwi93,Shi93,Har96,Tit98}.

It is common to treat the input state for optical Bell inequality
measurements as the singlet state
\begin{equation}
  \label{intro:singlet}
  |\psi_{\text{singlet}} \rangle
  = \Bigl( \vert + \rangle\otimes\vert - \rangle
        - \vert - \rangle\otimes\vert + \rangle \Bigr)
        / \sqrt{2} \, ,
\end{equation}
corresponding to an entanglement of vertical ($+$) and horizontal
($-$) polarized photons in a net zero--angular--momentum state.
However, PDC is not a perfect source of pairs of photons; one must
account for the higher--order contributions due to more than two
photons.  Also, the time of emission of the correlated pair is random.
These features of PDC will be shown to be accommodated in the group
theoretic approach of applying an SU(1,1) transformation to the vacuum
state~\cite{Tru85,Yur86}.

The Bell inequality test is performed, first by producing the photon
pairs via PDC, and then directing the photons through passive optical
elements (beam splitters, phase shifters, polarizer rotators).  These
passive optical elements mix two bosonic fields at each stage and
conserve photon number; such transformations are described as SU(2)
transformations~\cite{Tru85,Yur86,Cam89}.  Thus, the input vacuum state is
subjected to an overall unitary transformation which can be decomposed
into a sequence of SU(1,1) and SU(2) transformations, to produce the
final output state.  This state is then subjected to photon
coincidence measurements, and the constraints of local realism impose
an upper bound on photon coincidence rates for various parameter
choices.  A violation of this upper bound corresponds to a violation
of Bell's inequality and, hence, a test of local realism.

We shall see that it is natural to characterize Bell inequality
experiments in terms of unitary transformations and to identify the
Lie algebra which generates these transformations for particular Bell
inequality experiments.  We show that ideal Bell inequality
experiments effect an SU(1,1) transformation, which is distinct from
the SU(1,1) transformation that produces the photon pairs.  Distinct
ideal Bell inequality experiments can be identified with different
SU(1,1) subgroups in Sp(8,$\mathbb R$).

In Section~\ref{sec:background}, we describe tests of Bell's
inequality and establish the mathematical framework necessary for
studying such tests.  The SU(2) transformations for passive optical
elements and the SU(1,1) transformations for PDC are discussed.  In
Section~\ref{sec:Ideal}, we treat the ideal test of a Bell inequality
by analyzing the experimental arrangement of an SU(1,1) PDC
transformation followed by SU(2) passive optical elements.  The result
is that the ideal Bell inequality test arises as an SU(1,1) $\subset$
Sp(8,$\Bbb R$) transformation of the vacuum state with some freedom to
choose the applicable SU(1,1) transformation.  An alternative
realization of an ideal Bell inequality test is presented in
Section~\ref{sec:realizations} as well as an example of a
post--selected form of testing Bell's inequality.  Conclusions are
presented in Section~\ref{sec:Conclusions} and include a brief
discussion of the nature of the general SU(1,1) coherent state
involved in Bell inequality tests.

\section{Background}
\label{sec:background}

\subsection{The Bell inequality test}

In the standard Bell inequality test, a source produces a pair of
entangled spin-$1/2$ particles.  These two particles propagate in
different directions and are detected by spatially separated detectors
which can measure the spin state of each of the two particles along
specified axes.  An example of an entangled state is given by
Eq.~(\ref{intro:singlet}).  We refer to the two spatially separated
components (channels) as $a$ and $b$, and the state may be subjected
to simultaneous measurements of the spin states of $a$ and $b$ along
preferred axes.

The CHSH inequality version of the Bell inequality~\cite{Cla69}
introduces the figure of merit
\begin{equation}
  \label{eq:CHSH}
  S = \bigl\vert C(\theta_a,\theta_b) + C(\theta_a,\theta_b^\prime) +
    C(\theta_a^\prime,\theta_b)
  - C(\theta_a^\prime, \theta_b^\prime) \bigr\vert,
\end{equation}
with $\theta_a,\theta_a^\prime$ describing measurement axes for
system~$a$, $\theta_b,\theta_b^\prime$ for system~$b$ and
$C(\theta_a,\theta_b)$ the correlation between $a$ and $b$ (with
values in the range $[-1,+1]$).  Local realism places a bound of $2$
on $S$, giving the CHSH inequality,
\begin{equation}
  \label{eq:CHCHInequality}
  S \leq 2 \quad \text{(for local realism),}
\end{equation}
and quantum mechanics predicts a violation of this inequality for
certain quantum states~\cite{Bra92}.
For example, using the singlet state (\ref{intro:singlet}) with the
values~\cite{Wal94}
\begin{equation}
  \label{eq:ValuesForViolation}
  \theta_a - \theta_b = \theta'_a - \theta_b = \theta'_a - \theta'_b =
  \tfrac{1}{3}(\theta_a - \theta'_b) = \pi/8 \, , 
\end{equation}
one obtains a violation of the CHSH inequality of $S = 2\sqrt{2}$.

Here we employ the CHSH inequality to investigate Bell inequality
tests as unitary transformations.  A detailed analysis of
Bell inequalities requires consideration of the Clauser--Horne
formulation of the inequality~\cite{Cla74} and treatment of loopholes
in the various experimental tests~\cite{San91}.  However, these issues
are not directly relevant to this analysis, and the CHSH inequality
suffices to consider an ideal bound on a system which is governed by
local realism.

\subsection{The algebra sp(8,$\mathbb R$) and its subalgebras}

Although the Bell inequality test was devised using two spin-$1/2$
particles, we may use a boson representation to realize an optical
version of the experiment.  In this case there are four boson field
modes to consider, each with a corresponding annihilation operator:
$\hat{a}_+$ corresponding to the vertical polarization for the $a$
spatial mode, $\hat{a}_-$ corresponding to the horizontal polarization
for the $a$ spatial mode, and annihilation operators~$\hat{b}_\pm$ for
the vertical and horizontal polarizations for the $b$~spatial modes.
There are thus four mutually--commuting boson--operator pairs
$\hat{a}_{\pm}$, $\hat{b}_{\pm}$ and their conjugates, which can be
presented as
\begin{gather}
  \label{combining:c}
  \hat{a}_+ \to \hat{c}_1\, ,\quad \hat{a}_-\to \hat{c}_2 \, ,\quad 
  \hat{b}_+\to \hat{c}_3\, ,\quad \hat{b}_-\to \hat{c}_4 \, , \\
  \label{combining:cdagger}
  \hat{a}^\dagger_+\to \hat{c}^\dagger_1 \, ,\quad 
  \hat{a}^\dagger_-\to \hat{c}^\dagger_2 \, ,\quad
  \hat{b}^\dagger_+\to \hat{c}^\dagger_3 \, ,\quad 
  \hat{b}^\dagger_-\to \hat{c}^\dagger_4~.
\end{gather}
These operators obey the usual boson commutation rules
\begin{equation}
  \label{mathematical:quadratic}
  [\hat{c}_i,\hat{c}_j^{\dagger}]=\delta_{ij}\, ,\quad
  [\hat{c}_i,\hat{c}_j]=[\hat{c}_i^{\dagger},\hat{c}_j^{\dagger}]=0 \, .
\end{equation}

An optical test of Bell's inequality can employ parametric
down--conversion (PDC), polarization rotation (where the spin--$1/2$
state corresponds to a polarization state of the photon), beam
splitters, phase shifters and mirrors as stages of the processing of
the quantum state.  Each of these stages can be represented
mathematically as a unitary transformation provided that losses are
neglected.  The infinitesimal generators of these transformations
consist of quadratic combinations of the operators~(\ref{combining:c})
and (\ref{combining:cdagger}), of the form $\hat{c}_i \hat{c}_j$,
$\hat{c}_i^\dagger \hat{c}_j$ and $\hat{c}_i^\dagger
\hat{c}_j^\dagger$.  These quadratic operators span the
complexification of the algebra sp(8,$\mathbb R$), with the standard
basis ($i,j \in \{1,2,3,4\}$)
\begin{align}
  \label{A:noncompact}
  \hat{A}_{ij} &= \hat{c}^{\dagger}_{i} \hat{c}^{\dagger}_{j} \, , \\
  \label{C:compact}
  \hat{C}_{ij}&=\tfrac{1}{2}
  (\hat{c}^{\dagger}_{i}\hat{c}_{j}+\hat{c}_{j}\hat{c}^{\dagger}_i) \,
  , \\
  \label{B:noncompact}
  \hat{B}_{ij}&=\hat{c}_{i}\hat{c}_{j} \, .
\end{align}
These operators obey the (complexified) sp(8,${\mathbb R}$) commutation 
relations
\begin{align}
  \label{eq:sp(4,R)CommutationRelations}
  [\hat{A}_{ij},\hat{A}_{kl}] &= 0 =
  [\hat{B}_{ij},\hat{B}_{kl}]\, , \nonumber \\
  [\hat{C}_{ij},\hat{C}_{kl}] &=
  \delta_{jk}\hat{C}_{il}-\delta_{il}\hat{C}_{kj}\, , \nonumber \\
  [\hat{C}_{ij},\hat{A}_{kl}] &=
  \delta_{jk}\hat{A}_{il}+\delta_{jl}\hat{A}_{ik}\, , \\
  [\hat{C}_{ij},\hat{B}_{kl}] &=
  -\delta_{il}\hat{B}_{jk}-\delta_{ik}\hat{B}_{jl}\, ,  \nonumber \\
  [\hat{A}_{ij},\hat{B}_{kl}] &=
  -\delta_{ki}\hat{C}_{jl}-\delta_{kj}\hat{C}_{il} 
  -\delta_{il}\hat{C}_{jk}-\delta_{jl}\hat{C}_{ik}\, .\nonumber
\end{align}
Note that the generators $\{ \hat{C}_{ij} \}$ span a complex u(4)
subalgebra.  This four--boson realization of the algebra
sp(8,${\mathbb R}$) provides the language with which to describe Bell
inequality experiments (and many other optical experiments as well).

In optical versions of Bell inequality tests, the measurement of the
coincidence rate~$C(\theta_a,\theta_b)$ used in Eq.~(\ref{eq:CHSH})
for photons is to record simultaneous photodetections in spatial
modes~$a$ and~$b$.  A convenient expression for the correlation
function in terms of this four--boson realization is~\cite{Rei86}
\begin{equation}
  \label{coincidence}
  C(\theta_a,\theta_b) = \frac{ \bigl\langle ( a_+^\dagger a_+
        - a_-^\dagger a_- )
  ( b_+^\dagger b_+ - b_-^\dagger b_- ) \bigr\rangle }
  { \bigl\langle ( a_+^\dagger a_+ + a_-^\dagger a_- )
  ( b_+^\dagger b_+ + b_-^\dagger b_- ) \bigr\rangle } \, .
\end{equation}
Strictly speaking, this expression is applicable to the CHSH inequality
when the photon pair flux is sufficiently low that the probability of
more than one pair of photons arriving at the detectors is negligible.
The spontaneous generation of pairs by PDC permits a sufficiently
short interval to be chosen, in principle, to ensure that
higher--order terms (beyond the vacuum and photon pairs) can be
neglected.  The vacuum produces no coincidences and the coincidence rate
is set to zero in this case.  The normalization is trivial for the
case of a single pair, with photons arriving at~$a$ and~$b$ detectors.
The coincidence rate represented by~(\ref{coincidence}) is appropriate
for quantum optics experiments.  We show in Section~\ref{sec:Ideal}
that the flux rate of photon pairs cancels via the denominator, and,
therefore, the flux rate does not appear in calculations of Bell's
inequality.

The algebra sp(8,${\mathbb R}$) contains many subalgebras which have
physical significance in terms of quantum optics and Bell inequality
tests.  In the following, we identify certain subalgebras with optical
transformations induced by beam splitters, phase shifters,
polarization rotations, and PDCs.

\subsection{Realizations of su(2) subalgebras}
\label{sub:su2}

Many passive (i.e., photon number conserving) optical transformations
can be described by various su(2) subalgebras in sp(8,${\mathbb R}$).
For example, many useful su(2) subalgebras can be realized as a
two--boson realization for any $i \neq j$, given by
\begin{align}
  \hat{J}_x^{(ij)} 
  &= \tfrac{1}{2} (\hat{c}_i^\dagger \hat{c}_j + \hat{c}_i
  \hat{c}_j^\dagger)\, , \nonumber \\
  \hat{J}_y^{(ij)} 
  &= \tfrac{1}{2\text{i}} (\hat{c}_i^\dagger \hat{c}_j - \hat{c}_i
  \hat{c}_j^\dagger)\, , \\ 
  \hat{J}_z^{(ij)} 
  &= \tfrac{1}{2} ( \hat{c}_i^\dagger \hat{c}_i -
  \hat{c}_j^\dagger \hat{c}_j )\, , \nonumber
\end{align}
and satisfying $[\hat{J}_x^{(ij)},\hat{J}_y^{(ij)}] = \text{i}
\hat{J}_z^{(ij)}$ with $x$, $y$, $z$ cyclic.

Some of the realizations of these su(2) subalgebras correspond to
\begin{itemize}
\item mixing of two modes (interactions of the type $\hat{a}_+^\dagger
  \hat{b}_+ + \hat{a}_+ \hat{b}_+^\dagger$) via a beam splitter,
\item mixing polarizations in one mode ($\hat{a}_+^\dagger \hat{a}_- +
  \hat{a}_+ \hat{a}_-^\dagger$), and
\item mixing both spatial modes and polarization modes
  ($\hat{a}_\pm^\dagger \hat{b}_\mp$).
\end{itemize}

Consider, for example, the polarization--independent beam
splitter~\cite{Cam89}.  The generator associated to this optical
device is
\begin{align}
  \label{eq:BeamSplitterGenerator}
  \hat{J}_{\text{BS}} &= \hat{J}_x^{(13)} + \hat{J}_x^{(24)} 
  \nonumber \\ 
  &= \tfrac{1}{2}(\hat{a}_+ \hat{b}_+^\dagger +
  \hat{a}_+^\dagger \hat{b}_+ + \hat{a}_- \hat{b}_-^\dagger +
  \hat{a}_-^\dagger \hat{b}_-) \, .
\end{align}
The associated unitary transformation of a $50/50$
polarization--independent beam splitter is
\begin{equation}
  \label{eq:BeamSplitterTransformation}
  U_{\text{BS}} = \exp \bigl(\text{i}(\pi/4)\hat{J}_{\text{BS}} \bigr)
  \, ,
\end{equation}
which is an element of the SU(2) subgroup corresponding to
polarization--independent channel mixing~\cite{Hats}.  As another
example, the operator
\begin{align}
  \label{eq:PhaseShifterGenerator}
  \hat{J}_{\text{PS}} &= \hat{J}_z^{(13)} + \hat{J}_z^{(24)}
  \nonumber \\ 
  &= \tfrac{1}{2}(\hat{a}_+^\dagger \hat{a}_+ -
  \hat{b}_+^\dagger \hat{b}_+ + 
  \hat{a}_-^\dagger \hat{a}_- - \hat{b}_-^\dagger \hat{b}_- )
\end{align}
describes a polarization--independent phase shifter and also
generates a transformation in this same SU(2) subgroup.

As an example of mixing polarizations in one spatial mode, consider
the operator
\begin{equation}
  \label{eq:PolarizationRotatorGenerator}
  \hat{J}_a = \hat{J}_x^{(12)} 
  = \tfrac{1}{2}( \hat{a}_+^\dagger \hat{a}_- + \hat{a}_+
  \hat{a}_-^\dagger ) \, .
\end{equation}
This operator generates the unitary transformation
\begin{equation}
  \label{eq:PolarizationRotatorTransformation}
  U_a (\theta_a) = \exp \bigl( \text{i}\theta_a \hat{J}_a \bigr) \, ,
\end{equation}
which rotates the polarization in channel $a$ by an angle $\theta_a$
and does not affect channel $b$; i.e., this transformation describes a
polarization rotator of angle $\theta_a$ in the $a$ channel.

The above are just some of the su(2) subalgebras used to describe
lossless, passive optical elements: elements for which the total
number of input quanta equals the total number of output quanta.

\subsection{Realizations of su(1,1) subalgebras}
\label{sub:su11}

The transformations associated with parametric down--conversion (PDC)
are active; they create or annihilate pairs of photons.  The Lie
algebra su(1,1) has been shown~\cite{Tru85,Yur86} to describe these
transformations.

In PDC, a crystal with a $\chi^{(2)}$ nonlinearity is pumped by a
coherent field, wherein each pump photon spontaneously decays into a
pair of photons.  In {\em degenerate} PDC, the two photons in the pair
are identical; in {\em nondegenerate} PDC, the pump photon decays into
two non--identical photons.  For $(\mathbf{k}_i,\omega_i)$ the
wave vector and angular frequency of the $i^{\text{th}}$ field, with
$i=0$ for the pump field and $i=1,2$ for the two output fields, energy
conservation yields $\omega_0=\omega_1+\omega_2$, and momentum
conservation yields $\mathbf{k}_0 = \mathbf{k}_1 + \mathbf{k}_2$.  For
degenerate PDC, $\omega_1=\omega_2$ and $\mathbf{k}_1 =
\mathbf{k}_2$~\cite{Har96}.

For below--threshold operation, the pump field may be considered to be
a classical field.  Treating the pump field as classical allows the
annihilation and creation operators for the pump field photon to be
treated as c--numbers and not as operators.

\subsubsection{PDC and the algebra su(1,1)}
\label{subsub:PDC&su(1,1)}

By analogy with the beam splitter, which is described by an SU(2)
transformation, PDC is described by an SU(1,1) transformation.  
A basis for the su(1,1) algebra is given by the set of operators $\{
\hat{K}_x, \hat{K}_y, \hat{K}_z \}$, with commutation relations
\begin{equation}
  \label{K:algebra}
  [\hat{K}_x,\hat{K}_y]     =-\text{i}  \hat{K}_z, \;
  [\hat{K}_y,\hat{K}_z]     = \text{i}  \hat{K}_x, \;
  [\hat{K}_z,\hat{K}_x]     = \text{i}  \hat{K}_y \, .
\end{equation}

For degenerate PDC, the appropriate realizations of su(1,1) are
one--boson realizations given by the generators
\begin{align}
  \label{Kdegen:generators}
  \hat{K}_x^{(i)} &= \tfrac{1}{4} (\hat{c}_i^\dagger \hat{c}_i^\dagger 
  + \hat{c}_i \hat{c}_i )\, , \nonumber \\ 
  \hat{K}_y^{(i)} &= \tfrac{1}{4\text{i}} (\hat{c}_i^\dagger
  \hat{c}_i^\dagger - \hat{c}_i \hat{c}_i )\, , \\ 
  \hat{K}_z^{(i)} &= \tfrac{1}{4}  (\hat{c}_i^\dagger \hat{c}_i + \hat{c}_i
  \hat{c}_i^\dagger) \, . \nonumber 
\end{align}
Here the annihilation operator~$\hat{c}_i$ can refer to any
of~$\hat{a}_+$, $\hat{a}_-$, $\hat{b}_+$ or $\hat{b}_-$.

For the nondegenerate case, where the PDC generates two non--identical
photons, the appropriate realizations of su(1,1) are two--boson
realizations given by the generators
\begin{align}
  \label{Knondegen:generators}
  \hat{K}_x^{(ij)} &= \tfrac{1}{2}(\hat{c}_i^\dagger \hat{c}_j^\dagger 
  + \hat{c}_i \hat{c}_j)\, , \nonumber \\
  \hat{K}_y^{(ij)} &= \tfrac{1}{2\text{i}}(\hat{c}_i^\dagger
  \hat{c}_j^\dagger 
  - \hat{c}_i \hat{c}_j )\, , \\ 
  \hat{K}_z^{(ij)} &= \tfrac{1}{2}(\hat{c}_i^\dagger \hat{c}_i + \hat{c}_j
  \hat{c}_j^\dagger)\, . 
  \nonumber 
\end{align}
A type--I PDC is one for which, typically, $\hat c_i=\hat a_+$ and
$\hat c_j=\hat b_+$, whereas, in a type--II PDC, $\hat c_i=\hat a_+$
and $\hat c_i=\hat b_-$.  That is, a pair of photons is created in the
same polarization in type--I PDC, and a pair of photons is created in
opposite polarizations in type--II down conversion.

It is also possible to design PDCs which generate entangled
pairs~\cite{Kwi95}.  Such a setup involves a type--II PDC where the
emission directions of the $a$ and $b$ channel photons are made to
overlap, and is described by a four--boson realization of
su(1,1)~\cite{Bam95}.  There are several such realizations, each of
which describes the generation of a different entangled state.  One
example is given by the generators
\begin{align}
  \label{Kentangled:generators}
  \hat{K}_x &= \tfrac{1}{2}(\hat{a}_+^\dagger \hat{b}_-^\dagger -
  \hat{a}_-^\dagger \hat{b}_+^\dagger 
  + \hat{a}_+ \hat{b}_- -\hat{a}_- \hat{b}_+ )\, , \nonumber \\
  \hat{K}_y &= \tfrac{1}{2\text{i}}(\hat{a}_+^\dagger \hat{b}_-^\dagger -
  \hat{a}_-^\dagger \hat{b}_+^\dagger 
  - \hat{a}_+ \hat{b}_- +\hat{a}_- \hat{b}_+ )\, , \\
  \hat{K}_z &= \tfrac{1}{2}(\hat{a}_+^\dagger \hat{a}_+ + \hat{b}_-
  \hat{b}_-^\dagger
  + \hat{a}_-^\dagger \hat{a}_- + \hat{b}_+ \hat{b}_+^\dagger)\, .
  \nonumber 
\end{align}
It is interesting to note that the above four--boson realization is
a direct sum of two of the two--boson realizations in
Eq.~(\ref{Knondegen:generators}), with a sign change, as follows
\begin{align} 
  \hat{K}_x &= \hat{K}_x^{(14)}-\hat{K}_x^{(23)} \, , \nonumber \\
  \hat{K}_y &= \hat{K}_y^{(14)}-\hat{K}_y^{(23)} \, , \\
  \hat{K}_z &= \hat{K}_z^{(14)}+\hat{K}_z^{(23)} \, . \nonumber
\end{align}
One can easily check that these generators also satisfy the
commutation relations of su(1,1).  Adjusting the parameters of the PDC
(such as the relative phase) can lead to other similar four--boson
realizations.  It will be shown in the following that this particular
realization generates an SU(1,1) transformation which describes the
generation of the singlet state.

It is possible to design a different PDC that is also described by
this four--boson realization~\cite{Har96}, and which also generates
entangled pairs.  This setup, however, entangles the photons in
wave number rather than polarization.  Pairs of photons are selected by
four pinholes in a diaphragm placed downstream from the PDC to produce
four channels, labelled $1$ through $4$, with wave vectors $\{
\mathbf{k}_1, \mathbf{k}_2, \mathbf{k}_3, \mathbf{k}_4 \} $.  These
wave vectors satisfy
\begin{align}
  \label{eq:FourBosonWavevectorRelations}
  |\mathbf{k}_1| &= |\mathbf{k}_4|\, , \nonumber \\
  |\mathbf{k}_2| &= |\mathbf{k}_3|\, , \quad \text{but}\
  |\mathbf{k}_1| \neq |\mathbf{k}_2|\, , 
\end{align}
and
\begin{equation}
  \label{eq:FourBosonWavevectorRelations2}
  \mathbf{k}_1 + \mathbf{k}_3 = \mathbf{k}_2 + \mathbf{k}_4 =
  \mathbf{k} \, ,
\end{equation}
where $\mathbf{k}$ is the wave vector of the beam incident on the
crystal.

Let the annihilation operators for these four wave numbers correspond
to the ordered set $ \{ \hat{a}_+, \hat{b}_+, \hat{b}_-, \hat{a}_- \}
$.  Thus, we are able to employ the earlier notation although the
physical system is entirely different.  The su(1,1) algebra describing
this PDC is also given by the four--boson realization of
Eq.~(\ref{Kentangled:generators}).

\subsubsection{Pair generation using PDC}
\label{subsub:pairgeneration}

The rate of pair creation using PDC is proportional to the
nonlinearity~$\chi^{(2)}$, the strength of the (classical) pump field
and the interaction time.  In the following, we develop a
one--parameter transformation which describes pair generation from the
vacuum state for PDC.

Consider the action of an SU(1,1) transformation, generated by the
realization corresponding to either the degenerate PDC algebra of
Eq.~(\ref{Kdegen:generators}) or the nondegenerate PDC algebra of
Eq.~(\ref{Knondegen:generators}), on the vacuum state $|0\rangle$.
This state is an eigenstate of $\hat{K}_z$ and is annihilated by
$\hat{K}_- = \hat{K}_x - \text{i}\hat{K}_y$, and thus by using a
normal--ordered form, it is sufficient to express a general SU(1,1)
transformation of the vacuum state as the one--parameter
transformation
\begin{equation}
  \label{eq:PDCtransformation}
  \Upsilon(\gamma)|0\rangle = \exp \bigl( \text{i} \gamma \hat{K}_x \bigr)
  |0\rangle \, , \quad \gamma \in {\mathbb R} \, .
\end{equation}
The resultant state is not simply a pair of photons, but a
superposition of photon number states which also includes the vacuum,
pairs--of--pairs, and higher order contributions.  For $\gamma$ small,
the resulting state can be approximated as
\begin{equation}
  \label{PI:withvac}
  \Upsilon(\gamma)|0\rangle \approx |0\rangle + \text{i} \gamma
    \hat{K}_x |0\rangle \, . 
\end{equation}
The role of the vacuum in the superposition~(\ref{PI:withvac}) is to
include in the state the feature that the creation of the desired
photon pair occurs at a random time.  That is, the photon pair cannot
be created `on demand'.  Note that photon counting does not detect the
vacuum, so the inclusion of this state does not alter the final
measurement process.

Consider, for example, the case of a type--I nondegenerate PDC,
described by $\hat{K}_x^{(13)} = \tfrac{1}{2}(\hat{a}_+^\dagger
\hat{b}_+^\dagger + \hat{a}_+ \hat{b}_+)$.  The resulting
(approximate) state is
\begin{equation}
  \label{PI:withvacTypeI}
  \Upsilon_{\text{type-I}}(\gamma)|0\rangle 
  \approx |0\rangle + \tfrac{\text{i}}{2} \gamma |1,0,1,0 \rangle  \, ,
\end{equation}
where the Fock notation $|i,j,k,l\rangle$ describes a state with $i$,
$j$, $k$, $l$ photons in the $a_+$, $a_-$, $b_+$, $b_-$ modes,
respectively.

Provided that observations are conditioned on actually detecting
photons, the vacuum state plays no role.  Thus, `conditioning' the
state can be described by a projection $\hat{\pi}$, defined by
\begin{align}
  \label{eq:Projection}
  \hat{\pi} &= |1,0,1,0\rangle \langle 1,0,1,0| \nonumber \\
  &\qquad + |1,0,0,1\rangle \langle 1,0,0,1| \nonumber \\
  &\qquad + |0,1,1,0\rangle \langle 0,1,1,0| \\
  &\qquad + |0,1,0,1\rangle \langle 0,1,0,1|\, , \nonumber
\end{align}
which projects any state in the Fock space into the subspace of states
with exactly one photon in channel $a$ and one in channel $b$.  Such a
projection relies on photodetectors that can discriminate between one
and more than one photon~\cite{Kim99}.  As the photon--pair flux rate
is assumed to be small, contributions due to higher--order terms are
negligible, and thus current photodetectors which do not discriminate
between one and more photons are adequate.

By projecting the state in Eq.~(\ref{PI:withvacTypeI}), we obtain the
`conditioned photon pair state' $\vert 1,0,1,0 \rangle$.  We thus have
the requisite pair of correlated particles but not an entangled state.

A relevant basis for the subspace of degenerate
eigenstates of $\hat{\pi}$ is the so--called `Bell state basis', given
by
\begin{align}
  \label{eq:BellStateBasis}
  |\psi_\pm \rangle &= \tfrac{1}{\sqrt{2}} \bigl( |1,0,0,1\rangle \pm
   |0,1,1,0\rangle \bigr) \, , \nonumber  \\
  |\phi_\pm \rangle &= \tfrac{1}{\sqrt{2}} \bigl(|1,0,1,0\rangle \pm
  |0,1,0,1\rangle \bigr) \, .
\end{align}
The state $|\psi_- \rangle$ is the singlet state
$|\psi_{\text{singlet}} \rangle$ of Eq.~(\ref{intro:singlet}).  The
other Bell states are equally suitable entangled states for testing
the Bell inequality.  It is desirable, in tests of the Bell
inequality, to be able to generate entangled states such as these.

As an example of a realization of SU(1,1) that will generate an
entangled (Bell) state, consider the four--boson realization given by
Eq.~(\ref{Kentangled:generators}).  To lowest order in $\gamma$, we
have
\begin{align}
  \label{entpair}
  \Upsilon_{\text{singlet}}(\gamma) \vert 0\rangle
        &= \exp \bigl( \text{i} \gamma \hat{K}_x \bigr) \vert 0 \rangle
        \nonumber \\
        &\approx \vert 0 \rangle + \tfrac{\text{i}}{2}\gamma
        \bigl( |1,0,0,1 \rangle - | 0,1,1,0 \rangle \bigr) \, .
\end{align}
By applying the projection $\hat{\pi}$, the state reduces,
`conditioned' on photons being present, to the singlet state
$|\psi_-\rangle$.  In the experimental setup of Kwiat \emph{et
  al}~\cite{Kwi95}, a PDC described by this transformation has been
shown to generate the singlet state.  By adjusting the parameters of
the PDC and performing local unitary transformations, any of the Bell
states of Eq.~(\ref{eq:BellStateBasis}) can be produced; the
generators correspond to different four--boson realizations of su(1,1)
similar to that of Eq.~(\ref{Kentangled:generators}).  It is also
possible to use PDC to generate an entangled state in wave number,
using the method described above, and described by the same SU(1,1)
transformation.

\section{Realizations of the Bell inequality test}
\label{sec:realizations}

\subsection{The ideal Bell inequality test}
\label{sec:Ideal}

In this section we construct simple transformations on the vacuum
state which correspond to an ideal Bell inequality experiment.  
We establish the algebra which generates these transformations to be
su(1,1), and calculate the quantum--mechanical correlation functions
$C(\theta_a,\theta_b)$ for the corresponding state.

\begin{figure}
  \includegraphics*[width=3.25in,keepaspectratio]{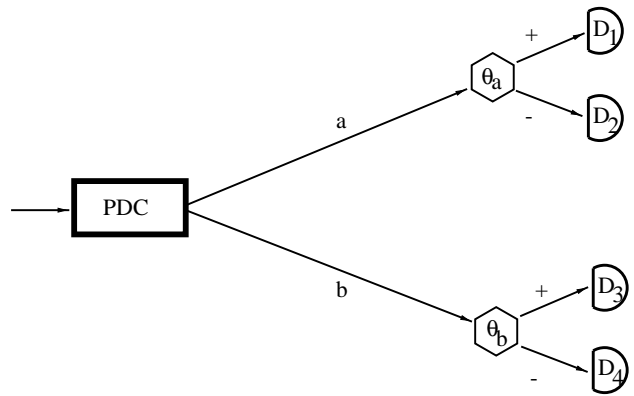}
  \caption{Diagrammatic representation of the ideal Bell inequality
    experiment.  The PDC which produces the singlet state is used.
    Channel $a$ is passed through a polarization analyzer at angle
    $\theta_a$, and channel $b$ through one at angle $\theta_b$.
    Photodetectors $D_1$, $D_2$, $D_3$, and $D_4$ measure the
    corresponding photocounts.}
  \label{fig:Ideal}
\end{figure}

The ideal Bell inequality experiment is depicted in
Fig.~\ref{fig:Ideal}.
This experiment has been performed by Kwiat \emph{et al}~\cite{Kwi95}.
The first requirement is a device which produces the entangled photon
pairs, thereby providing the necessary Bell state.  The singlet state
of Eq..~(\ref{entpair}) is obtained by using the projector $\hat \pi$
on a state produced via PDC described by the four--boson realization
of Eq.~(\ref{Kentangled:generators}).  The presence of the vacuum
state in the superposition signifies that the `location', or `creation
time', of the singlet is indeterminate; singlet states are not created
`on demand'.  The SU(1,1) transformation
$\Upsilon_{\text{singlet}}(\gamma)$ is generated by the operator
\begin{equation}
  \label{four:kp}
  \hat{K} = \tfrac{1}{2}(\hat{a}_+^\dagger \hat{b}_-^\dagger -
  \hat{a}_-^\dagger \hat{b}_+^\dagger 
  + \hat{a}_+ \hat{b}_- -\hat{a}_- \hat{b}_+ ) \, .
\end{equation}
Note that the Bell inequality test can be performed with any one of
the four Bell states in Eq.~(\ref{eq:BellStateBasis});
our choice of the singlet state is simply for convention.  Thus, this
choice of $\hat{K}$ as the Bell state generator is not unique.

The PDC output is directed to local polarization rotators, one for the
$a$ mode and one for the $b$ mode, each followed by a polarizing beam
splitter.  The polarizing beam splitter separates the two orthogonal
polarization components of the field and directs them to two
photodetectors, which can count the photons in each of the two
polarizations.  We refer to the combination of the polarizer rotator,
with an adjustable parameter $\theta_{a,b}$, and the polarizing beam
splitter, which separates the two polarizations into distinct spatial
modes, as a polarization analyzer.  This polarization analyzer is
depicted as the hexagon in Fig.~\ref{fig:Ideal}.

Bell's inequality establishes an upper bound to the measurable photon
coincidence rate allowed by local realistic assumptions for various
choices of $\theta_a$ and $\theta_b$ of the two polarization analyzers.
The polarizations are transformed independently by a U(1)$_a \otimes
$U(1)$_b$ rotation, with two independent, local parameters~$\theta_a$
and~$\theta_b$, with the following two mutually--commuting generators.
For polarization rotation of the $a$ mode, the generator
$\hat{J}_a$ of Eq.~(\ref{eq:PolarizationRotatorGenerator}) is required; 
similarly, for the $b$~mode, we require
\begin{equation}
  \label{subgroups:polb}
  \hat{J}_b = \hat{J}_x^{(34)} = \tfrac{1}{2}(\hat{b}_+^\dagger
  \hat{b}_- + \hat{b}_+ \hat{b}_-^\dagger) \, . 
\end{equation}
Equal polarization rotations for modes $a$ and $b$ leave the
singlet state invariant, as
\begin{equation}
  \label{eq:EqualRotationsCommute}
  [\hat{K}, \hat{J}_a + \hat{J}_b] = 0\, .
\end{equation}
Thus, it is only necessary~\cite{nonlocal} to consider a difference
transformation $U_-(\theta_-)$ generated by $\hat{J} = \hat{J}_a -
\hat{J}_b$, given by
\begin{equation}
  \label{eq:PolarizationDifferenceRotation}
  U_-(\theta_-) = e^{\text{i} \theta_- \hat{J}_a} e^{-\text{i}
  \theta_- \hat{J}_b} = e^{\text{i} \theta_- \hat{J}}\, .
\end{equation} 
Note that the operators $\hat{J}$, $\hat{K}$, and
\begin{equation}
  \label{eq:operatorLy}
  \hat{L} = \tfrac{1}{2\text{i}}( \hat{a}_-^\dagger \hat{b}_-^\dagger -
  \hat{a}_+^\dagger \hat{b}_+^\dagger 
  - \hat{a}_- \hat{b}_- + \hat{a}_+ \hat{b}_+ )\, ,
\end{equation}
close under commutation to form a realization of su(1,1):
\begin{equation}
  \label{eq:IdealBellAlgebra}
  [\hat{J},\hat{K}] = \text{i} \hat{L}, \quad 
  [\hat{L},\hat{J}] = \text{i} \hat{K}, \quad 
  [\hat{K},\hat{L}]= -\text{i} \hat{J} \, .
\end{equation}
This realization of su(1,1) is distinct from any of the realizations
describing PDC.  This algebra generates the Lie group SU(1,1), which
can be applied to the ground state to generate the state
\begin{align}
\label{ideal:U}
  | \gamma, \theta_- \rangle &= U_- (\theta_-)
  \Upsilon_{\text{singlet}}(\gamma)|0\rangle \nonumber \\
  &= e^{\text{i} \theta_- \hat{J} } e^{ \text{i}\gamma \hat{K} }|0\rangle \, .
\end{align}
The transformation of the ground state $|0\rangle$ consists of a PDC
transformation to generate an entangled state, followed by local
polarization rotations on the $a$ and $b$ modes by angles $\theta_-$
and~$-\theta_-$, respectively.  To lowest order, the state
(\ref{ideal:U}) is a superposition of a vacuum state and a two--photon
state.  Neglecting the vacuum state, the effective state is then a
Bell state (\ref{eq:BellStateBasis}) if $\theta_-=0$.  However, for
general~$\theta_-$, the two--photon contribution to the superposition
is an entanglement of non--orthogonal SU(2) coherent
states~\cite{Wan00}.

The correlation function $C(\theta_a,\theta_b)$ of
Eq.~(\ref{coincidence}) for the state $|\gamma,\theta_a -
\theta_b\rangle$ is given by
\begin{equation}
  \label{eq:ExpectationC(a,b)}
  C(\theta_a,\theta_b) = \frac{\langle \gamma, \theta_a - \theta_b |
  (\hat{\sigma}_z)_a (\hat{\sigma}_z)_b | 
  \gamma, \theta_a - \theta_b \rangle}
  {\langle \gamma, \theta_a - \theta_b |
  (\hat{\sigma}_0)_a (\hat{\sigma}_0)_b | 
  \gamma, \theta_a - \theta_b \rangle} \, ,
\end{equation}
where $(\hat{\sigma}_z)_a = \hat{a}_+^\dagger \hat{a}_+ -
\hat{a}_-^\dagger \hat{a}_-$, $(\hat{\sigma}_0)_a = \hat{a}_+^\dagger
\hat{a}_+ + \hat{a}_-^\dagger \hat{a}_-$, and likewise for
$(\hat{\sigma}_z)_b$ and $(\hat{\sigma}_0)_b$.

Note that, by using the approximation for $\gamma$ small of
Eq.~(\ref{entpair}) and `conditioning' the state on photons being
present (i.e., excluding the vacuum state), the PDC generates the
singlet state $|\psi_- \rangle$ of Eq.~(\ref{eq:BellStateBasis}).
Calculating the correlation function for this state, one obtains the
familiar result
\begin{align}
  \label{eq:C(a,b)Approx}
  C(\theta_a, & \theta_b) \nonumber \\
  &= \bigl\langle \psi_{-} \bigl|
  U_-^{-1}(\theta_a - \theta_b) 
  [(\hat{\sigma}_z)_a (\hat{\sigma}_z)_b]
  U_-(\theta_a-\theta_b)
  \bigr| \psi_{-} \bigr\rangle \nonumber \\
  &= - \cos 2(\theta_a - \theta_b)\, .
\end{align}
The singlet state can lead to a violation $S=2\sqrt{2}$ for the
parameter choices (\ref{eq:ValuesForViolation}).

There is an interesting su(1,1) structure to the correlation function
$C(\theta_a,\theta_b)$, which we detail as follows.
Beginning with the numerator, we first obtain the result
\begin{equation}
  \label{eq:ExactStep1}
  \begin{split}
  U_-^{-1}&(\theta_a - \theta_b)[(\hat{\sigma}_z)_a
  (\hat{\sigma}_z)_b] U_-(\theta_a - \theta_b) \\
  &= \Bigl( \cos(\theta_a - \theta_b) (\hat{\sigma}_z)_a
  - \sin(\theta_a - \theta_b) (\hat{\sigma}_y)_a \Bigr)  \\
  &\qquad \cdot \Bigl(
  \cos(\theta_a - \theta_b) (\hat{\sigma}_z)_b
  + \sin(\theta_a - \theta_b) (\hat{\sigma}_y)_b \Bigr) \, .
  \end{split}
\end{equation}
Then, consider the following change of basis:
\begin{align}
  \label{eq:ExactStep2}
  \hat{J}_z^+ &= (\hat{\sigma}_z)_a + (\hat{\sigma}_z)_b \, , &\quad
  \hat{J}_z^- &= (\hat{\sigma}_z)_a - 
  (\hat{\sigma}_z)_b \, , \nonumber \\
  \hat{J}_y^+ &= (\hat{\sigma}_y)_a + (\hat{\sigma}_y)_b \, , &\quad
  \hat{J}_y^- &= (\hat{\sigma}_y)_a - (\hat{\sigma}_y)_b \, .
\end{align}
This basis transforms simply under the action of
$\Upsilon_{\text{singlet}}(\gamma)$ as follows:
\begin{align}
  \label{eq:ExactStep3}
  \Upsilon_{\text{singlet}}^{-1}(\gamma) \hat{J}_z^+
  \Upsilon_{\text{singlet}}(\gamma) &= \hat{J}_z^+ \, , \nonumber \\
  \Upsilon_{\text{singlet}}^{-1}(\gamma) \hat{J}_y^+
  \Upsilon_{\text{singlet}}(\gamma) &= \hat{J}_y^+ \, , \\
  \Upsilon_{\text{singlet}}^{-1}(\gamma) \hat{J}_z^-
  \Upsilon_{\text{singlet}}(\gamma) &=
  \cosh (\gamma) \hat{J}_z^- + \sinh (\gamma) \hat{L}_z \, , \nonumber \\
  \Upsilon_{\text{singlet}}^{-1}(\gamma) \hat{J}_y^-
  \Upsilon_{\text{singlet}}(\gamma) &=
  \cosh (\gamma) \hat{J}_y^- + \sinh (\gamma) \hat{L}_y \, , \nonumber
\end{align}
where
\begin{align}
  \label{eq:ExactStep4}
  \hat{L}_z &= \tfrac{1}{2\text{i}} (\hat{a}_+^\dagger \hat{b}_-^\dagger +
  \hat{a}_-^\dagger 
  \hat{b}_+^\dagger - \hat{a}_+ \hat{b}_- + \hat{a}_- \hat{b}_+) \, ,
  \nonumber \\ 
  \hat{L}_y &= \tfrac{1}{2} (\hat{a}_-^\dagger \hat{b}_-^\dagger +
  \hat{a}_+^\dagger 
  \hat{b}_+^\dagger + \hat{a}_- \hat{b}_- + \hat{a}_+ \hat{b}_+) \, .
\end{align}
Evaluating the numerator of Eq.~(\ref{eq:ExpectationC(a,b)}) gives
\begin{align}
  \label{eq:ExactStep5}
  \langle \gamma, \theta_a - \theta_b |
  (\hat{\sigma}_z)_a &(\hat{\sigma}_z)_b | 
  \gamma, \theta_a - \theta_b \rangle \nonumber \\
  &= \sinh^2 (\gamma) \Bigl( -\cos^2 (\theta_a - \theta_b) \langle 0 |
  \hat{L}_z^2 | 0 \rangle \nonumber \\
  &\qquad + \sin^2 (\theta_a - \theta_b) \langle 0 |
  \hat{L}_y^2 | 0 \rangle \Bigr) \nonumber \\
  &= - \tfrac{1}{2} \sinh^2 (\gamma) \cos 2(\theta_a - \theta_b) \, .
\end{align}
We have utilized the fact that the mixed terms $\sin
(\theta_a-\theta_b) \cos (\theta_a-\theta_b) (\hat{\sigma}_z)_a
(\hat{\sigma}_y)_b$, etc., in Eq.~(\ref{eq:ExactStep1}) do not contribute,
and also that the vacuum expectation values for the operators of the
`J-type' (of the form $\hat{c}_i^\dagger \hat{c}_j$) vanish.

Next, evaluating the denominator in a similar fashion, we first
observe that
\begin{equation}
  \label{eq:ExactStep6}
  U_-^{-1}(\theta_a - \theta_b)[(\hat{\sigma}_0)_a 
  (\hat{\sigma}_0)_b] U_-(\theta_a - \theta_b)
  = (\hat{\sigma}_0)_a (\hat{\sigma}_0)_b \, .
\end{equation}
Again, consider the change of basis
\begin{equation}
  \label{eq:ExactStep7}
  \hat{N}_0^+ = (\hat{\sigma}_0)_a + (\hat{\sigma}_0)_b \, , \quad
  \hat{N}_0^- = (\hat{\sigma}_0)_a - (\hat{\sigma}_0)_b \, .
\end{equation}
This basis transforms simply under the action of
$\Upsilon_{\text{singlet}}(\gamma)$ as follows:
\begin{align}
  \label{eq:ExactStep8}
  \Upsilon_{\text{singlet}}^{-1}(\gamma) \hat{N}_0^+
  \Upsilon_{\text{singlet}}(\gamma) &=
  \cosh (\gamma) \hat{N}_0^+ + \sinh (\gamma) \hat{L}_0 \, ,\nonumber \\
  \Upsilon_{\text{singlet}}^{-1}(\gamma) \hat{N}_0^-
  \Upsilon_{\text{singlet}}(\gamma) &= \hat{N}_0^- \, ,
\end{align}
where
\begin{equation}
  \label{eq:ExactStep9}
  \hat{L}_0 = -\tfrac{1}{2\text{i}} (\hat{a}_+^\dagger \hat{b}_-^\dagger -
  \hat{a}_-^\dagger 
  \hat{b}_+^\dagger - \hat{a}_+ \hat{b}_- + \hat{a}_- \hat{b}_+) \, .
\end{equation}
Evaluating the denominator gives
\begin{align}
  \label{eq:ExactStep10}
  \langle \gamma, \theta_a - \theta_b |
  (\hat{\sigma}_0)_a (\hat{\sigma}_0)_b | 
  \gamma, \theta_a - \theta_b \rangle &=
  \sinh^2 (\gamma) \langle 0 | \hat{L}_0^2 | 0 \rangle \nonumber \\
  &= \tfrac{1}{2} \sinh^2 (\gamma) \, .
\end{align}
Thus, we find that the correlation $C(\theta_a,\theta_b)$ is given by
\begin{equation}
  \label{eq:C(a,b)Exact}
  C(\theta_a,\theta_b) = - \cos 2(\theta_a - \theta_b) \, .
\end{equation}
The result is identical to the correlation function of the singlet
state, given by Eq.~(\ref{eq:C(a,b)Approx}), and is independent of
the flux rate term~$\gamma$.  The cancellation of~$\gamma$ occurs
because of the normalization with respect to the cross--correlation of
total number of photons at $a$ and $b$.  Although the dependence
on~$\gamma$ vanishes in the expression, it is assumed that~$\gamma$ is
sufficiently small to ensure that the probability of more than one pair
arriving at the detectors is negligible over the detector integration
time per event.

This simple formulation of the Bell inequality test reveals a basic
su(1,1) structure to the experiment.  In the following, this result is
shown to be general for several realized experiments.  There is
considerable choice of the su(1,1) $\subset$ sp(8,${\mathbb R}$)
subalgebra that can be used, depending on the type of Bell state
generated and the corresponding optical transformations performed on
it.  It will be shown in the following how existing experiments use
other su(1,1) subalgebras distinct from the $JKL$ algebra to test the
Bell inequality.

\subsection{Alternative ideal Bell inequality test}
\label{subsec:alternative}

\begin{figure}
  \includegraphics*[width=3.25in,keepaspectratio]{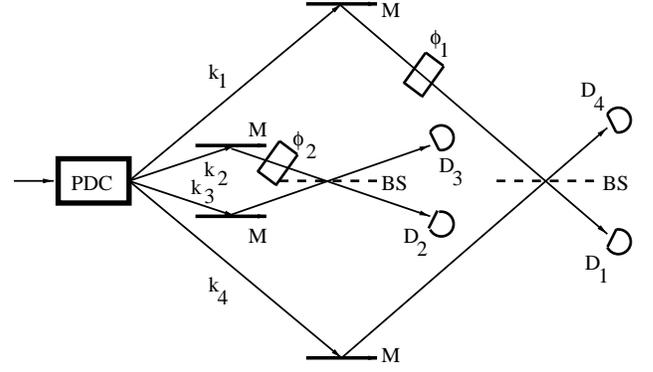}
  \caption{Schematic of the experiment of Horne \emph{et al} for
    testing the Bell inequality.}
  \label{fig:Horne}
\end{figure}

Although the ideal Bell inequality test has been presented in terms of
entangled photons with respect to polarization, an alternative test was
suggested by Horne {\em et al}~\cite{Hor89}, depicted in
Fig.~\ref{fig:Horne}, and realized experimentally by Rarity and
Tapster~\cite{Rar90}.
This realization employs a PDC which creates entanglement in
wave number rather than polarization, as described by a four--boson
realization of su(1,1) similar to that of
Eq.~(\ref{Kentangled:generators}).

Using the description of photon pairs entangled in wave number given in
Section~\ref{subsub:PDC&su(1,1)}, the appropriate generator for
producing entangled pairs is
\begin{equation}
  \label{eq:AlternativeKx}
  \hat{K}^\prime = \tfrac{1}{2}( \hat{a}_+^\dagger \hat{b}_-^\dagger +
  \hat{a}_-^\dagger \hat{b}_+^\dagger 
  + \hat{a}_+ \hat{b}_- + \hat{a}_- \hat{b}_+)\, ,
\end{equation}
which produces a pair of photons with wave numbers $k_1$ and $k_3$,
entangled with a pair of photons with wave numbers $k_2$ and $k_4$.
(The use of the `prime' on $\hat{K}^\prime$ is meant to distinguish
this generator from that of Eq.~(\ref{four:kp}).)
Employing the approximation that only one photon pair is created,
$\hat{K}^\prime$ generates the Bell state
$|\psi_+ \rangle$ of Eq.~(\ref{eq:BellStateBasis}).
Rather than subjecting these fields to polarization rotation, phase
shifts ($\phi_1, \phi_2$) are applied, and the corresponding
generators, following Eq.~(\ref{eq:PhaseShifterGenerator}) and the
notation of Section~\ref{sub:su2}, are
\begin{equation}
  \label{A0prime}
  \hat{J}_{\text{PS}}^a = \hat{J}_z^{(12)} = \tfrac{1}{2} ( \hat{a}_+^\dagger
  \hat{a}_+ - \hat{a}_-^\dagger \hat{a}_- ) \, , 
\end{equation}
and 
\begin{equation}
  \label{A3}
  \hat{J}_{\text{PS}}^b = \hat{J}_z^{(34)} = \tfrac{1}{2} ( \hat{b}_+^\dagger
  \hat{b}_+ - \hat{b}_-^\dagger \hat{b}_- ) \, . 
\end{equation}
Similar to the ideal case, only the phase shift \emph{difference}
between the two channels will actually transform the entangled state,
and thus we apply the generator
\begin{align}
  \label{eq:AlternativePhaseShift}
  \hat{J}^\prime &= \hat{J}_{\text{PS}}^a -
  \hat{J}_{\text{PS}}^b \, , \nonumber \\ 
  &= \tfrac{1}{2}( \hat{a}_+^\dagger \hat{a}_+ - \hat{a}_-^\dagger
  \hat{a}_- - \hat{b}_+^\dagger \hat{b}_+ + \hat{b}_-^\dagger
  \hat{b}_- ) \, ,
\end{align}
in the form of the unitary operator
\begin{equation}
  \label{eq:AlternativePhaseShiftTransformation}
  U_{\text{PS}} (\phi_-) = \exp \bigl( \text{i} \phi_- \hat{J}^\prime
  \bigr) \, , \quad \phi_- = \phi_1 - \phi_2 \, .
\end{equation}
Note that the entangled state generator $\hat{K}^\prime$, the phase shift
operator $\hat{J}^\prime$, and the operator
\begin{equation}
  \label{eq:AlternativeLy}
  \hat{L}^\prime = \tfrac{1}{2\text{i}}(\hat{a}_+^\dagger \hat{b}_-^\dagger 
  - \hat{a}_-^\dagger \hat{b}_+^\dagger - \hat{a}_+ \hat{b}_- +
  \hat{a}_- \hat{b}_+ )\, ,
\end{equation}
close to an su(1,1) algebra with commutation relations
\begin{equation}
  \label{eq:AlternativeAlgebra}
  [\hat{J}^\prime,\hat{K}^\prime] = \text{i} \hat{L}^\prime \, , \quad
  [\hat{L}^\prime,\hat{J}^\prime] = \text{i} \hat{K}^\prime \, , \quad 
  [\hat{K}^\prime,\hat{L}^\prime] =-\text{i} \hat{J}^\prime \, .
\end{equation}

The experimental scheme involves an interferometric arrangement for
the phase shifts to be meaningful; the fields must be mixed by a
wavelength--independent 50/50 beam splitter (BS), as described by
$U_{\text{BS}}$ of Eq.~(\ref{eq:BeamSplitterTransformation}).

The apparatus performs a transformation on the vacuum state to give
the entangled state
\begin{equation}
  \label{alternative:U}
  |\gamma, \phi_- \rangle 
        = U_{\text{BS}} \cdot e^{\text{i} \phi_- \hat{J}^\prime} \cdot 
        e^{ \text{i} \gamma \hat{K}^\prime } |0\rangle \, ,
\end{equation}
followed by photon coincidence detection in each of the four output
modes (detectors $D_i,\ i=1,2,3,4$).  Since the vacuum is invariant
under the transformation $U_{\text{BS}}$, we can express this
transformation as
\begin{align}
  \label{alternative:U2}
  |\gamma, \phi_- \rangle 
        &= U_{\text{BS}} \cdot e^{\text{i} \phi_- \hat{J}^\prime} \cdot
        e^{ \text{i} \gamma \hat{K}^\prime } \cdot U_{\text{BS}}^{-1}
        |0\rangle \, , \\
        &= e^{\text{i} \phi_- (U_{\text{BS}}\hat{J}^\prime
        U_{\text{BS}}^{-1})} \cdot
        e^{ \text{i} \gamma (U_{\text{BS}}\hat{K}^\prime
        U_{\text{BS}}^{-1}) } |0\rangle \, . \nonumber
\end{align}
Thus, the transformation on the vacuum can be expressed as a SU(1,1)
transformation generated by the algebra (\ref{eq:AlternativeAlgebra}),
{\it conjugated} by $U_{\text{BS}}$.  The relevant su(1,1)
subalgebra for this alternative Bell inequality test is spanned by the
operators $U_{\text{BS}}\hat{J}^\prime U_{\text{BS}}^{-1}$,
$U_{\text{BS}}\hat{K}^\prime U_{\text{BS}}^{-1}$, and
$U_{\text{BS}}\hat{L}^\prime U_{\text{BS}}^{-1}$.

Note that the generator $U_{\text{BS}}\hat{K}^\prime U_{\text{BS}}^{-1}$
can be calculated to be
\begin{equation}
  \label{eq:alternativeBSTransformedK}
  U_{\text{BS}}\hat{K}^\prime U_{\text{BS}}^{-1} = - \tfrac{1}{2}(
  \hat{a}_+^\dagger \hat{b}_+^\dagger - \hat{a}_-^\dagger
  \hat{b}_-^\dagger
  + \hat{a}_+ \hat{b}_+ - \hat{a}_- \hat{b}_-)\, ,
\end{equation}
and thus the approximate Bell state generated by this operator is
$|\phi_-\rangle$ of Eq.~(\ref{eq:BellStateBasis}).  Thus, the
experiment proposed by Horne \emph{et al} is equivalent to an ideal
Bell inequality test using the entangled Bell state $|\phi_-\rangle$.

\subsection{Post--selected Bell inequality test}
\label{subsec:postselected}

\begin{figure}
  \includegraphics*[width=3.25in,keepaspectratio]{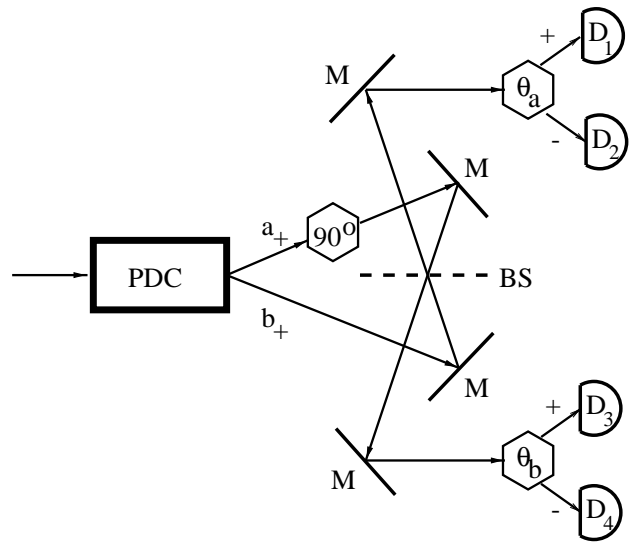}
  \caption{Schematic of the experiment of Ou and Mandel for
    testing the Bell inequality.}
  \label{fig:OuMandel}
\end{figure}

We have seen that the ideal Bell inequality experiment can be
described as an appropriate SU(1,1) transformation on the ground
state.  However, not all Bell inequality experiments are equivalent to
the ideal test given in Section~\ref{sec:Ideal}, yet nonetheless test
the Bell inequality.  A particularly salient example is the
post--selected Bell inequality test of Ou and Mandel~\cite{Ou88}.
Although the experiment was designed to test the Clauser--Horne
version of the Bell inequality~\cite{Cla74}, a simplified version of
the experimental arrangement, depicted in Fig.~\ref{fig:OuMandel},
would test the CHSH inequality and suffices for this analysis.

An important difference between this arrangement and those depicted in
Figs.~\ref{fig:Ideal} and~\ref{fig:Horne} is that, in this scheme, it
is as likely for both photons to go to analyzer~$a$ or~$b$ as having
one photon going to~$a$ and one photon to~$b$.  The projection of the
state produced in the scheme depicted in Fig.~\ref{fig:OuMandel} is a
post--selection process whereby the vacuum contribution is removed (no
detections occur), higher-pair contributions are neglected (such events
are rare) and the case that two photons go to the same region, $a$ or
$b$, is detected with photon counting detectors that discriminate
between one and two photons arriving.

In the absence of a photodetector which discriminates between the
arrival of one and two photons, the cases where both photons go to one
detector is registered as a single--photon detection.  This
single--photon detection is not distinguishable from a background of
single--photon events that arise due to detector inefficiencies, and,
therefore, photon--pair events arriving at one detector introduce a
loophole~\cite{San96}.  This problem may be rectified with new
detectors that do discriminate between one and two photons being
detected~\cite{Kim99}, and these detectors are being used for Bell
inequality tests in the Ou--Mandel scheme~\cite{Tak00}.

In the Ou--Mandel experiment, correlated photon pairs are generated by
a type--I PDC, described by a transformation of the form of
Eq.~(\ref{PI:withvac}), i.e., the transformation
\begin{equation}
  \label{eq:PostSelectedPDCTransformation}
  \Upsilon_{\text{OM}}(\gamma) = \exp \bigl( \text{i} \gamma
  \hat{K}_{\text{OM}} \bigr) \, ,
\end{equation}
where
\begin{equation}
  \label{eq:PostSelectedPDCGenerator}
  \hat{K}_{\text{OM}} = \hat{K}_x^{(13)} 
  = \tfrac{1}{2}(\hat{a}_+^\dagger
  \hat{b}_+^\dagger + \hat{a}_+ \hat{b}_+) \, .
\end{equation}
This transformation produces correlated photons but does not produce
an entangled pair.  To do so, the polarization of the $a$ port is
rotated by $90^\circ$, which is described by the transformation
$U_a(\theta_a)$ of Eq.~(\ref{eq:PolarizationRotatorTransformation})
with $\theta_a = \pi/2$, and entanglement is then produced by a
polarization--independent $50/50$ beam splitter (BS), described by the
transformation $U_{\text{BS}}$ of
Eq.~(\ref{eq:BeamSplitterTransformation}).  The result of all these
transformations on the vacuum state is to produce the state
\begin{align}
  \label{eq:PostSelectedEntangledState}
  |\psi(\gamma)\rangle &= U_{\text{BS}} U_a(\pi/2) 
        \Upsilon_{\text{OM}}(\gamma) |0\rangle \nonumber \\
  &= \bigl( U_{\text{BS}} U_a(\pi/2) \bigr) \Upsilon_{\text{OM}}(\gamma)
        \bigl( U_{\text{BS}} U_a(\pi/2) \bigr)^{-1} |0\rangle
        \nonumber \\
  &= \Upsilon_{\text{OM}}^\prime (\gamma) |0\rangle,
\end{align}
where we define $\Upsilon_{\text{OM}}^\prime(\gamma)$ to be the
conjugated transformation
\begin{align}
  \label{eq:PostSelectedUpsilonPrime}
  \Upsilon_{\text{OM}}^\prime (\gamma) &= \bigl( U_{\text{BS}} U_a(\pi/2) 
        \bigr) \Upsilon_{\text{OM}}(\gamma) 
        \bigl( U_{\text{BS}} U_a(\pi/2) \bigr)^{-1}
        \nonumber \\
  &= \exp \bigl( \text{i} \gamma \hat{K}_{\text{OM}}^\prime \bigr) \, ,
\end{align}
with
\begin{equation}
  \label{eq:PostSelectedPDCGeneratorPrime}
  \hat{K}_{\text{OM}}^\prime = \tfrac{1}{4}\Bigl( (\hat{a}_-^\dagger +
  \hat{b}_-^\dagger)
  (\hat{b}_+^\dagger - \hat{a}_+^\dagger) + (\hat{a}_- +
  \hat{b}_-)(\hat{b}_+ - \hat{a}_+) \Bigr).
\end{equation}

As mentioned above, the generator for entangled pair production
includes the possibility that both photons may go to polarizer $a$,
with none at $b$, and vice versa.  By expressing
$\hat{K}_{\text{OM}}^\prime$ as the sum $\hat{K}_{\text{OM}}^\prime =
\hat{K}_{\text{OM}}^1 + \hat{K}_{\text{OM}}^2$, with
\begin{align}
  \label{eq:DecompositionOfOMPairGenerator}
  \hat{K}_{\text{OM}}^1 &= \tfrac{1}{4}(\hat{a}_-^\dagger
  \hat{b}_+^\dagger - \hat{a}_+^\dagger 
  \hat{b}_-^\dagger + \hat{a}_- \hat{b}_+ - \hat{a}_+ \hat{b}_-)\, ,
  \nonumber \\ 
  \hat{K}_{\text{OM}}^2 &= \tfrac{1}{4}(\hat{b}_+^\dagger
  \hat{b}_-^\dagger - \hat{a}_+^\dagger 
  \hat{a}_-^\dagger + \hat{b}_+ \hat{b}_- - \hat{a}_+ \hat{a}_-)\, ,
\end{align}
it is clear that $\hat{K}_{\text{OM}}^1$ generates an entangled pair
(specifically, the singlet state), whereas $\hat{K}_{\text{OM}}^2$
generates photon pairs both travelling either to channel~$a$ or
channel~$b$.  The latter events cannot enable tests of local realism.
By post--selecting, the experiment essentially disregards the
component of the state generated by $\hat{K}_{\text{OM}}^2$ and
considers only the singlet component generated by
$\hat{K}_{\text{OM}}^1$.

We can state the idea of post--selection formally using the projection
$\hat{\pi}$ of Eq.~(\ref{eq:Projection}).  Projecting the state
$|\psi(\gamma)\rangle$ of Eq.~(\ref{eq:PostSelectedEntangledState})
gives
\begin{equation}
  \label{eq:PostSelectionProjection}
  \hat{\pi}\bigl(|\psi(\gamma)\rangle\bigr) \to |\psi_- \rangle \, ,
\end{equation}
where $|\psi_- \rangle$ of Eq.~(\ref{eq:BellStateBasis}) is the
singlet state.

The operator $\hat{J}_a$ of
Eq.~(\ref{eq:PolarizationRotatorGenerator}) describes the final
polarization rotation for the $a$--mode prior to photodetection, and
the corresponding $b$--mode operator is~$\hat{J}_b$ of
Eq.~(\ref{subgroups:polb}).  The transformations that these operators
perform on the state generated by
$\Upsilon^\prime_{\text{OM}}(\gamma)$ of
Eq.~(\ref{eq:PostSelectedUpsilonPrime}) are not trivial.  However, if
post--selection is performed (by applying the projection $\hat{\pi}$),
then the resulting transformations become identical to that of the
ideal Bell inequality test.

Thus, the Bell inequality test of Ou and Mandel is distinct from the
ideal test presented in Section~\ref{sec:Ideal}.  However, if
post--selection is given by the projection $\hat{\pi}$ of
Eq.~(\ref{eq:Projection}), then the test becomes equivalent to the
ideal test.  It should be noted again that the realization of this
projection relies on photodetectors which can distinguish between
different multiple photon events, such as two photons in channel $a$
and zero in channel $b$.

Whereas the disadvantage of the Ou--Mandel scheme is the need for
post--selection, an advantage is the relatively high flux of photon
pairs from type--I PDC compared to the production of
entangled--polarization pairs via PDC~\cite{Kwi95}.  For applications
of Bell state measurements to quantum teleportation and other schemes,
higher pair flux is an advantage.

\section{Conclusions}
\label{sec:Conclusions}

In studies of Bell inequalities, it is common to assume from the
outset that one is supplied with one of the four Bell
states~(\ref{eq:BellStateBasis}).  In quantum optical experiments,
such states are generated from the vacuum state by an SU(1,1)
transformation corresponding to parametric down conversion (PDC).
Local manipulations of the output state from the PDC are described by
SU(2) transformations.  Using these basic facts, we establish that
Bell inequality experiments, which manipulate four bosonic fields, are
SU(1,1) $\subset$ Sp(8,$\Bbb R$) transformations, and that distinct
four--boson realizations of SU(1,1) correspond to different
experiments.  For the post--selected Bell inequality, a projection
operator is necessary to recover the SU(1,1) transformation equivalent
to the ideal Bell inequality test.

This analysis is useful for a number of reasons.  It is useful to know
that an optical realization of the ideal Bell inequality test, which
begins with a vacuum state as a source, is described by a four--boson
realization of SU(1,1) to enable classification and comparison between
differing tests of Bell inequalities as well as to consider new tests.
According to the formalism we establish, new optical tests of Bell
inequalities would arise as distinct realizations of SU(1,1) $\subset$
Sp(8,$\Bbb R$).  The question of various distinct tests of local
realism can thus be related to the mathematical question of distinct
realizations of the subgroup SU(1,1) in Sp(8,$\Bbb R$) and the
transformations which relate these subgroups.  This question may be
relevant to continuous--variable approaches to tests of Bell
inequalities~\cite{Ral00} where degenerate PDC and the one--boson
realization~(\ref{Kdegen:generators}) are used.

The employment of a unitary description of Bell inequality tests is
useful as it includes higher--order photon number contributions and
incorporates the non--deterministic creation time for pairs of
photons. It also establishes a scheme for classifying existing
Bell inequality tests and proposing new tests.  In addition to the
importance of Bell inequalities, not only for testing local realism,
but also for their relevance to quantum cryptography~\cite{Eke91}, the
approach employed in this paper can be extended to studying quantum
teleportation~\cite{Ben93}, quantum dense coding~\cite{Ben92} and
entanglement swapping~\cite{Zuk93}.  These concepts and experiments in
quantum information apply the Bell states~(\ref{eq:BellStateBasis})
and their measurements to larger systems.  The application of group
theoretical methods to such systems follows from the analysis in this
paper and is under investigation.

Finally, the group theoretic approach establishes that the Bell
inequality apparatus, described as a unitary transformation, produces
an output state which can be regarded as a generalized coherent
state~\cite{Per86}; these coherent states are distinct from the
bi--pair (four--boson) coherent states investigated by Bambah and
Agarwal~\cite{Bam95}, as the relevant group is not a direct product
SU(1,1)$\otimes$SU(1,1).  The output coherent state is the transformed
vacuum state.  However, the vacuum state~$\vert 0 \rangle$ is not a
lowest weight state for the relevant realizations of SU(1,1) to
describe Bell inequality tests.  The representation containing the
vacuum state is certainly reducible.  However, the description of the
state as a generalized coherent state does provide a useful method for
thinking about the state which arrives at the photodetectors.  In this
way of thinking, the state may be characterized by probability
distributions for measurements, and the correlation
function~(\ref{eq:ExpectationC(a,b)}) can be regarded as being related
to a covariance of a joint probability distribution for
$(\hat\sigma_z)_a$ and~$(\hat\sigma_z)_b$ for this generalized
coherent state.  The elegance of the calculations in
Section~\ref{sec:Ideal} of the Bell inequality violations suggests
that there is something natural about considering these generalized
coherent states in such studies.

\begin{acknowledgments} 
  This research has been supported by a Macquarie University Research
  Grant and an Australian Research Council Large Grant.  DAR
  acknowledges the support of a Macquarie University Research
  Fellowship.  We acknowledge helpful discussions with M.\ 
  Pavi\v{c}i\'c, M.\ Revzen and M.\ \.Zukowski.  HdG would like to
  thank Marco Bertola for numerous discussions of the mathematical
  aspects of early versions of the work.
\end{acknowledgments}


\begin{thebibliography}{99}
\bibitem{Hats}
        We follow the convention of using a carat on
        operators (e.g.\ $\hat{J}$) and the symbol $U$ to represent a
        unitary transformation (group element) generated by such an
        operator.
\bibitem{nonlocal}
        This transformation is not local.  If one prefers
        to work with only local transformations, the group U(1)$_a \otimes
        $U(1)$_b$ can be applied, describing separate local
        transformations, and leads to the same results.
\end{thebibliography}
\end{document}